# Spectroscopic evidence for a first-order transition to the orbital Fulde-Ferrell-Larkin-Ovchinnikov state

Zongzheng Cao[1#], Menghan Liao[2#*], Hongyi Yan[3#], Yuying Zhu[4], Liguo Zhang[4], Kenji Watanabe[5], Takashi Taniguchi[6], Alberto F. Morpurgo[2], Haiwen Liu[3*], Qi-Kun Xue[1,4,7,8] and Ding Zhang[1,4,8,9*]

[1] State Key Laboratory of Low Dimensional Quantum Physics and Department of Physics, Tsinghua University, Beijing 100084, China

[2] Department of Quantum Matter Physics, University of Geneva, Geneva 1211, Switzerland

[3] Center for Advanced Quantum Studies, Department of Physics, Beijing Normal University, Beijing 100875, China

[4] Beijing Academy of Quantum Information Sciences, Beijing 100193, China

[5] Research Center for Electronic and Optical Materials, National Institute for Materials Science, 1-1 Namiki, Tsukuba 305-0044, Japan

[6] Research Center for Materials Nanoarchitectonics, National Institute for Materials Science, 1-1 Namiki, Tsukuba 305-0044, Japan

[7] Southern University of Science and Technology, Shenzhen 518055, China

[8] Frontier Science Center for Quantum Information, Beijing 100084, China

[9] RIKEN Center for Emergent Matter Science (CEMS), Wako, Saitama 351-0198, Japan

# These authors contributed equally.

*Email: menghan.liao@unige.ch
       haiwen.liu@bnu.edu.cn
       dingzhang@mail.tsinghua.edu.cn

# Abstract

A conventional superconducting state may be replaced by another dissipationless state hosting Cooper pairs with a finite momentum, leaving thermodynamic footprints for such a phase transition. Recently, a novel type of finite momentum pairing, so-called orbital Fulde-Ferrell-Larkin-Ovchinnikov (FFLO) state, has been proposed to occur in spin-orbit coupled superconductors such as bilayer 2H-NbSe$_2$. So far, a thermodynamic demonstration, which is key for establishing this exotic phase, has been lacking. Here, we reveal a first-order quantum phase transition to the orbital FFLO state in tunneling spectroscopic measurements on multilayer 2H-NbSe$_2$. The phase transition manifests itself as a sudden enhancement of the superconducting gap at an in-plane magnetic field $B_\parallel$ well below the upper critical field. Furthermore, this transition shows prominent hysteresis by sweeping $B_\parallel$ back and forth and quickly disappears once the magnetic field is tilted away from the sample plane by less than one degree. We obtain a comprehensive phase diagram for the orbital FFLO state and compare it with the theoretical calculation that takes into account the rearrangement of Josephson vortices. Our work elucidates the microscopic mechanism for the emergence of the orbital FFLO state.

Time reversal symmetry typically guarantees that singlet Cooper pairing with zero net momentum is energetically favored in conventional superconductors. Breaking time reversal symmetry changes the energy balance and Cooper pairing with a finite momentum becomes possible. One particular way to achieving this state is by pairing up electrons across the spin-split bands caused by the Zeeman effect in a small section of the Fermi surface[1-4]. This so-called Fulde-Ferrell-Larkin-Ovchinnikov (FFLO) state can possess enhanced upper critical magnetic fields at low temperatures, which is important for applications. However, realizing the FFLO state requires a clean superconductor—a longer mean free path than the coherence length—and a large Maki parameter—a much larger orbital pair-breaking magnetic field than the paramagnetic field. These demanding conditions limit its existence to a handful of materials.

Recently, an alternative to the standard (Zeeman-type) FFLO state has emerged[5-7]. It takes into account the modulation of transport momenta by a magnetic vector potential. This latter type of finite momentum pairing is dubbed the orbital FFLO state. Candidate materials are non-centrosymmetric superconducting multilayers with strong spin-orbit coupling. In a $NbSe_2$ bilayer, for example, inversion symmetry breaking in each Se-Nb-Se layer induces Ising pairing that quenches the spin degree of freedom[8-14]. Applying an in-plane magnetic field to this system therefore mainly induces opposite momentum shifts of Cooper pairs in the top and bottom layers. The formation of such an orbital FFLO state can give rise to an up-turn in the temperature dependence of the upper critical field, similar to the standard FFLO but at a temperature much closer to the transition temperature ($T_c$).

The up-turn feature has been reported in thin flakes of $NbSe_2$ (about 20 nm thick)[6] and lithium intercalated $MoS_2$ bilayers[7]. Other characteristics that are consistent with the orbital FFLO state, such as the six-fold symmetry, have also been observed in $NbSe_2$[6]. However, a complete demonstration of the FFLO state requires the detection of a phase transition and establishing its first-order nature, which are still missing. The small thickness of the candidate systems (NbSe$_2$ thin flakes or Li$_x$MoS$_2$ bilayers) coupled with mesoscopic sample sizes poses a serious challenge for applying the standard detection schemes, such as heat capacity and torque magnetometry[15-20], thermal conductance[21], or nuclear magnetic resonance[22-26]. Apart from experimental verification, the theoretical understanding of the orbital FFLO state in a multi-layer system is also limited. Under an in-plane magnetic field, NbSe$_2$ with a thickness of about 20 nm apparently hosts an anisotropic vortex matter—the so-called Josephson vortex[27-29]. A lattice of Josephson vortices may emerge such that its deformation and low-energy excitation can dominate the thermodynamic processes[30]. Its competition with the orbital FFLO state remains unexplored.

Here, we study the orbital FFLO state in thin flakes of NbSe$_2$ through a so-far unexplored route—tunneling spectroscopy—and report solid evidence for the first-order phase transition. First, by ramping up the in-plane magnetic field, the superconducting gap shows a sudden enhancement at a field well below the upper critical field. Secondly, the increase of the superconducting gap is hysteretic with the sweeping field and shows discrete jumps in the transition region. Finally, tilting the magnetic field slightly away (about 1 degree) from the in-plane direction quickly wipes out the above-mentioned first-order transition. These experimental features fit well with our theory that considers the competition between a Josephson vortex lattice and a state of Cooper pairs with layer-dependent transport momenta.

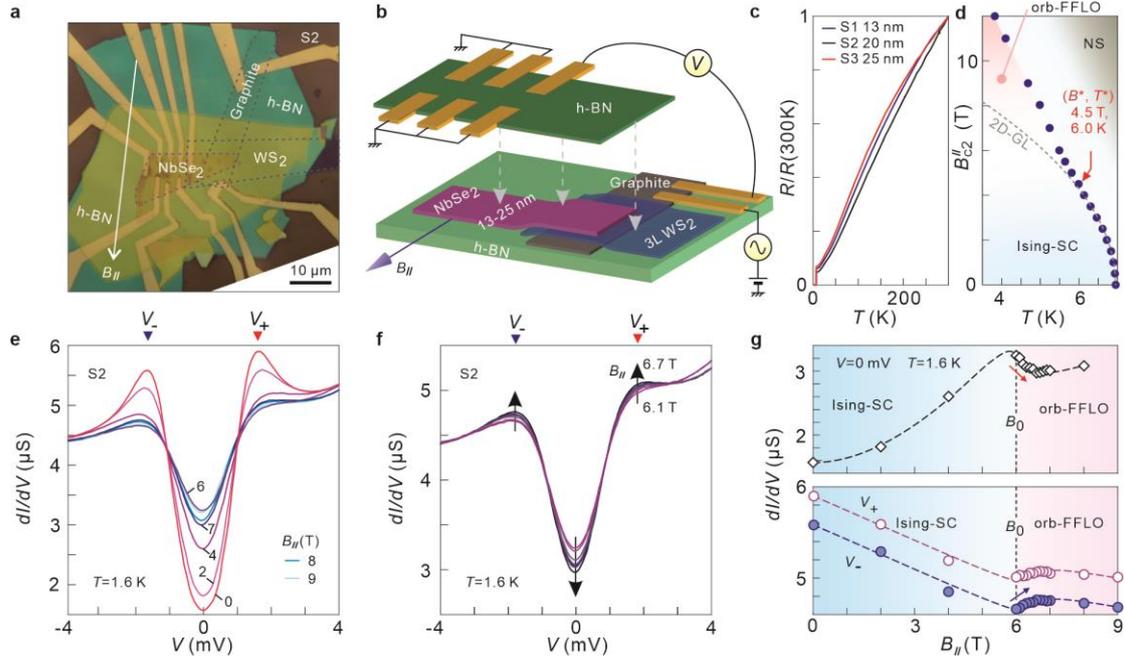

***Figure 1 Tunneling spectroscopy on NbSe₂.*** *a Optical image of sample S2. b Schematic drawing of the experimental setup in an exploded view. c Normalized resistances of samples S1-S3 as a function of temperature. d Temperature dependence of the in-plane upper critical magnetic field of sample S2. Filled symbols are data points by using the criterion of 50% normal state resistance. Dashed curve is a theoretical fit by using the 2D Ginzburg-Landau (2D-GL) formula. Colored shades indicate three phases: orbital FFLO state (orb-FFLO), Ising superconducting state (Ising-SC), and normal state (NS). Red arrow marks the crossover point where the data points start to deviate from the 2G-GL formula. e, f Tunneling spectra of sample S2 at 1.6 K and under different in-plane magnetic fields ($B_\parallel$). Red and blue triangles indicate the positions of two superconducting coherence peaks. Black arrows in panel f indicate the enhanced gap features with increasing $B_\parallel$. g Tunneling conductance at zero bias (top) or at the positions of the superconducting coherent peaks (bottom). Curves are guide to the eye. $B_0$ marks the critical magnetic field for the transition.*

Our experimental setup combines resistance characterizations and tunneling spectroscopy. Figure 1a is an optical image of sample S2. A thin flake of NbSe₂, together with the tunneling probe—a graphite bottom electrode and a trilayer WS₂ tunnel barrier, are encapsulated between two hexagonal boron nitride (h-BN) layers. Electrical contacts to NbSe₂ are realized by metallization of the trenches dry etched in the top h-BN layer (Fig. 1b). The h-BN encapsulation guarantees high quality of NbSe₂. Figure 1c shows that the residual resistivity ratios (RRR) for sample S1, S2 and S3 are 17,22 and 15, with the corresponding $T_c$ around 7 K. Consistent with the former report[6], our sample possesses an up-turn in the in-plane upper critical magnetic field, which deviates from a simple square root behavior. Figure 1d showcases the data from sample S2. Here, $B_{c2,\parallel}$ deviates from a simple 2D Ginzburg-Landau fitting. We define a point where this deviation starts as $T^* = 0.86\, T_c$ and $B^* = 4.5$ T (indicated by the arrow). Previous resistance measurements[6,7] suggested that the orbital FFLO state would occur when $T < T^*$ and $B_\parallel > B^*$. We will show that $(B^*, T^*)$ is only an approximate indicator whereas the tunneling spectroscopy determines the exact phase boundary between the Ising superconducting state and the orbital FFLO state. Figure 1e and f collect the tunneling spectra for sample S2 (20 nm thick) under different $B_\parallel$ at 1.6 K. At $B_\parallel = 0$ T, the tunneling spectrum shows strongly suppressed conductance around zero bias with peaks at around $\pm 1.6$ mV—typical features of a superconducting gap[31-33]. The normalized spectrum can be nicely described by the Dynes formula.

By increasing $B_\parallel$, the superconducting gap gets strongly suppressed but surprisingly not in a monotonic fashion. Figure 1e and f show that the conductance minimum around zero bias becomes shallower from 0 to 6 T but deepens from 6 to 7 T. This is further followed by a continuous suppression at higher $B_\parallel$ (Fig. 1e). Figure 1g summarizes the in-plane field dependences of the tunneling conductance at zero bias (top) and at two finite bias voltages where the superconducting coherence peaks situate (bottom). At an onset magnetic field $B_0$ of about 6 T, the zero-bias conductance suddenly drops (indicated by the arrow) and the conductance at the coherent peaks enhances. They suggest an abrupt enhancement in the superconducting gap $\Delta$. The value of $B_0$ is much smaller than $B_{c2,\parallel}$ obtained from resistance measurements at the same temperature (Fig. 1d). It marks a striking transition within the superconducting state. Furthermore, $B_0$ is above the empirical onset field $B^*$ for the orbital FFLO state. We therefore conclude that $B_0$ is the critical $B_\parallel$ where the sample transitions from Ising pairing to the orbital FFLO state.

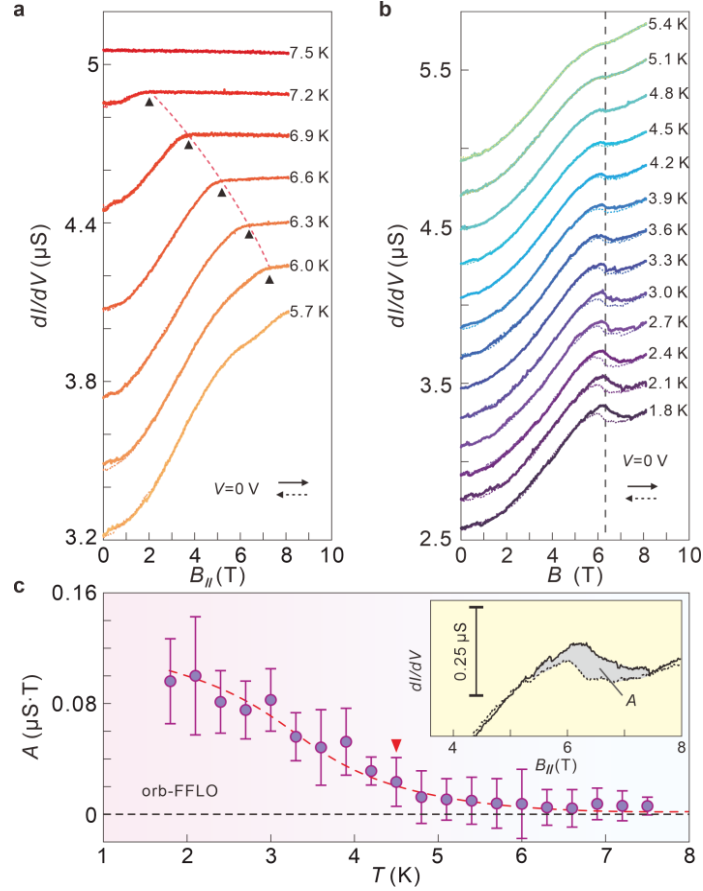

***Figure 2 Temperature dependence of zero-bias tunneling conductance (ZBTC). a, b** ZBTC as a function of in-plane magnetic field at a series of temperature points. Data are from sample S2. Solid and dotted curves are obtained from positive and negative sweeps. Black arrows in panel **a** denote the critical magnetic fields where the gap closes. Dashed line in panel **b** marks the local maximum of the curve at 1.8 K in the positive sweep. **c** Area encircled by the hysteresis loop of ZBTC from 5 to 8 T. Error bar represents the measurement noise. To evaluate the noise, we first sum up the difference between the positive and negative sweeps from 0 to 4 T at a fixed temperature. This value is then rescaled by taking into account the different ranges used for this summation and that for estimating the hysteresis loop area. Red arrow indicates the onset temperature for the hysteresis. Inset illustrates the hysteresis loop area.*

Figure 2 presents further evidence for the phase transition. Here, we continuously monitor the zero-bias tunneling conductance (ZBTC) as a function of $B_\parallel$ at various temperatures below $T_c$ (except for curves at 7.2 and 7.5 K). At relatively high temperatures (6 to 7 K), the ZBTC continuously increases with $B_\parallel$ before saturating at high magnetic fields (Fig. 2a). It reflects a second order phase transition from the superconducting state to the normal state[34]. At a lower temperature and a slightly higher field than the crossover point ($B^*, T^*$), the second order transition gets disrupted and a kink in ZBTC starts to develop (Fig. 2b). Further lowering the temperatures results in a kink at around 6 T in the otherwise smooth curve. This kink evolves to show pronounced hysteresis between the two traces obtained in opposite sweep directions (solid and dotted). This hysteretic behavior clearly demonstrates the first-order nature of a phase transition. We sum up the area enclosed by the hysteretic loop and plot it as a function of temperature in Fig. 2c. It suggests an onset temperature of 4.5 K for the hysteresis.

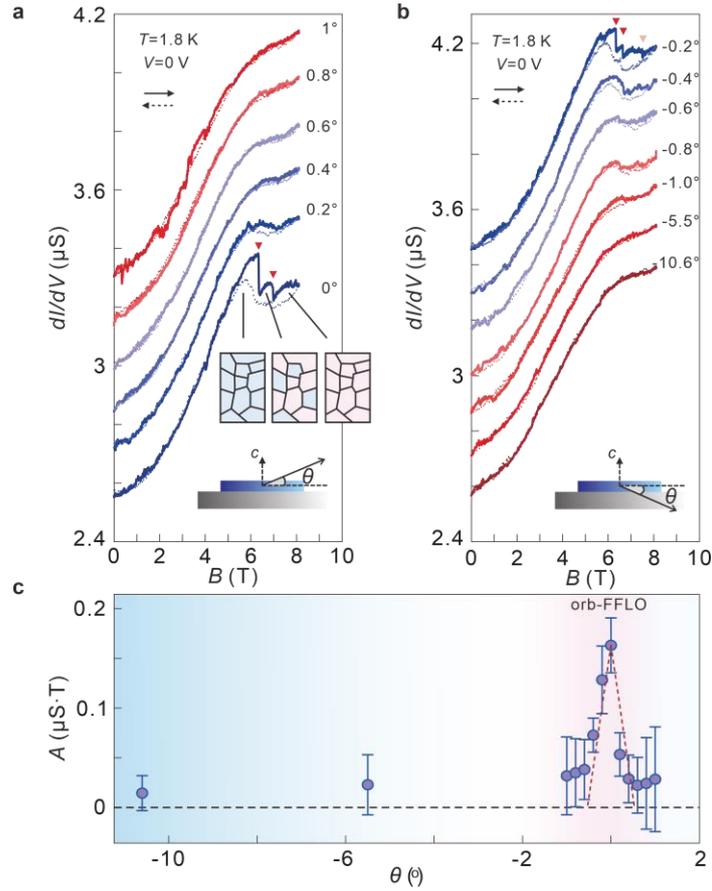

*Figure 3 Tilt-field dependence of zero-bias tunneling conductance (ZBTC). a, b* ZBTC as a function of the total magnetic field at a fixed temperature for different tilting angles (schematically shown in the insets). Data are from sample S2. Solid and dotted curves are obtained from positive and negative sweeps. Red arrows indicate multiple jumps. Inset in **a** schematically illustrates the phase transition in multiple domains. Blue and pink areas depict domains for the Ising superconductivity and the orbital FFLO state, respectively. **c** Tilting dependence of the hysteretic region in ZBTC around 6 T (from 5 to 8 T). Error bar represents the measurement noise, which is evaluated by the same protocol described in the caption of Fig. 2.

One defining character of the orbital FFLO state is its sensitive dependence on the direction of the magnetic field relative to the crystalline axis[6]. Whereas $B_\parallel$ in the *a-b* plane stabilizes the orbital FFLO by introducing the momentum shift of Cooper pairs, a small out-of-plane magnetic field along the *c*-axis disrupts its formation by introducing Abrikosov vortices. This fragility is readily seen in the tilt-field data in Fig. 3. We rotate the sample inside the solenoid magnet *in-situ* and measure the tunneling conductance as a function of the total magnetic field (Fig. 3a,b). Strong hysteresis with multiple jumps emerges at around 6 T when the magnetic field is well aligned to the in-plane direction (Fig. 3a). The multiple jumps are presumably due to the presence of several domains that transition to the orbital FFLO state sequentially with a varying magnetic field (as illustrated in the inset of Fig. 3a). This sawtooth behavior becomes less discernible at a tiny tilt angle $\theta$ of 0.4 degree. Further tilting to 1 degree removes the hysteresis. Figure 3c summarizes the hysteresis area as function of tilt angle, showing a rapid decay as $\theta$ deviates from 0 degree. (In fact, these analyses suggest that the data in Fig. 2 may have been taken with a slight misalignment of about 0.1°.)

Apart from sample S2, we carry out the same tunneling spectroscopy on two more samples (S1: 13 nm, $T_c = 6.74$ K; S3: 25 nm, $T_c = 7.31$ K). Figure 4a provides an overview of this thickness dependent study. From sample S2 to S3, the magnetic field for the first order transition shifts from 6.33 T to 4.62 T at comparable measurement temperatures $T \ll T_c$. Increasing the sample thickness weakens Ising superconductivity such that the phase boundary between Ising pairing and the orbital FFLO state moves to a lower magnetic field. For the thinnest sample (S1), on the other hand, we observe no phase transition down to 380 mK for $B_\parallel$ up to 12 T, although its resistance measurements reveal an upturn in $B_{c2,\parallel}$ at $B^* = 11$ T (4.2 K). This mismatch is probably because $B_0$—the critical field for the phase transition—is larger than $B^*$. Thinner samples may also experience stronger disorder effect which suppresses the formation of the orbital FFLO state.

In Fig. 4b, we plot the phase diagram of S2 and S3 by color-coding the ZBTC data

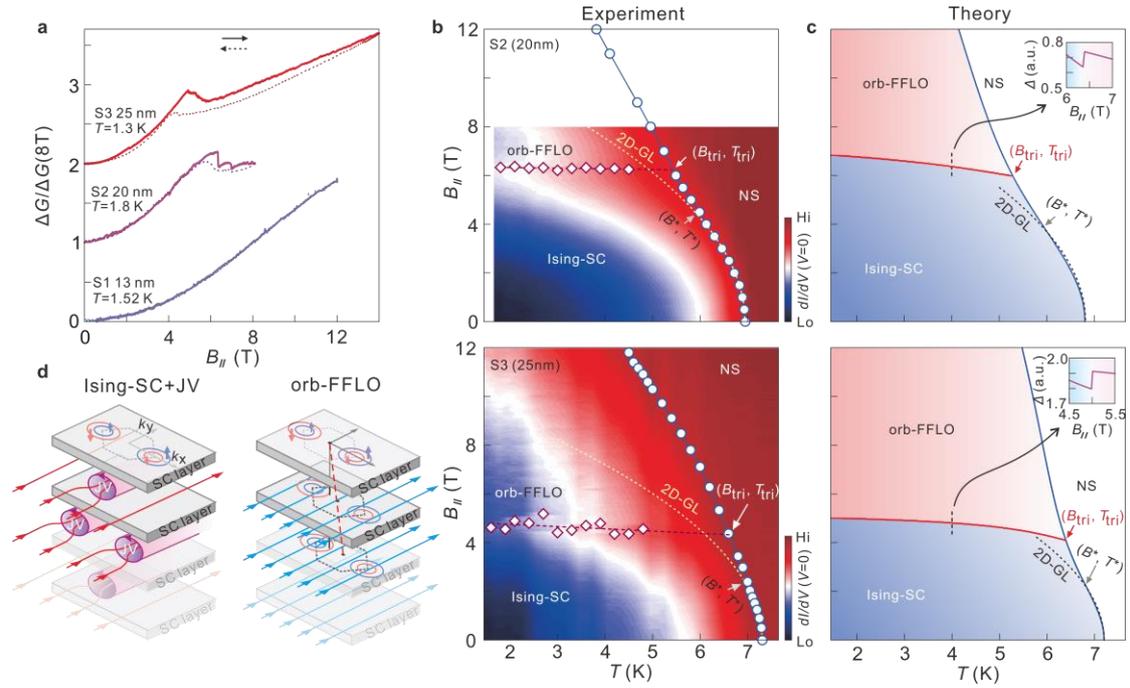

*Figure 4 Thickness dependence and the phase diagram*. **a** Summary of the normalized zero-bias tunneling conductance (ZBTC) as a function of in-plane magnetic field for three samples with different thicknesses. Here $\Delta G$ is ZBTC at a certain field subtracted by the zero-field value. Curves are vertically offset for clarity. **b** Color plot of ZBTC as a function of in-plane magnetic field and temperature. Here the data are obtained in the positive field sweeps. Diamonds are local maxima in the difference between ZBTC of positive and negative field sweeps. Dashed lines are linear fits. They represent the phase boundary between Ising superconducting state (Ising-SC) and the orbital FFLO state (orb-FFLO). Circles are $B_{c2,\parallel}$ obtained from resistance measurements. Dotted curves are fitting to the high temperature section of $B_{c2,\parallel}(T)$ by using the 2D Ginzburg-Landau (GL) formula. **c** Theoretically derived phase diagrams at two values of the Josephson coupling strength between the superconducting layers. Insets show the abrupt change of the superconductor order parameter. The thickness of the superconducting layer d and the inter-layer spacing D is chosen to be d/D = 2/3. **d** Schematic drawing of the distribution of magnetic fluxes in the states before and after the phase transition. On the left, we consider the Josephson vortices (JV) formed between the Ising superconducting (SC) layers. On the right, the flux lines distribute uniformly in the orb-FFLO state.

(obtained in the positive field sweep) as a function of $T$ and $B_\parallel$. The phase boundary between Ising superconductivity and the orbital FFLO state—$B_0(T)$—is mapped out from the local maxima (diamonds) in the difference between the ZBTC data in opposite sweeping directions of $B_\parallel$. In general, $B_0(T)$ slightly decreases with increasing temperature. In the relatively high temperature regime, thermodynamic fluctuations seem to smear out the first-order transition. Nevertheless, we linearly extrapolate $B_0(T)$ and it crosses with $B_{c2,\parallel}(T)$ (obtained from resistance measurements) at a point: $(B_{tri}, T_{tri})$—the tri-critical point for Ising superconductivity, the orbital FFLO state and the normal state. As shown in Fig. 4c, we theoretically reconstruct the same phase diagram. We employ the Lawrence-Doniach model to characterize NbSe$_2$ as a stack of Ising superconductors coupled by interlayer Josephson tunneling[29,35]. In agreement with experiment, the theoretically determined tri-critical point is also at a higher $B_\parallel$ and lower $T$ than the crossover point ($B^*$, $T^*$). In fact, the crossover point reflects the interference of Ising pairing within each Se-Nb-Se layer by the Josephson coupling. It serves only as an approximate indicator for the orbital FFLO state. We find that the phase boundary between Ising pairing and the orbital FFLO state is mainly determined by the interlayer Josephson coupling strength: a weaker interlayer Josephson coupling results in a smaller critical magnetic field for the phase transition.

The theoretical modelling helps reveal the microscopic picture of the first-order phase transition within the superconducting state, which we schematically draw in Fig. 4d. By ramping up the in-plane magnetic field, the Ising superconductor becomes interspersed with Josephson vortices. A Josephson vortex lattice can emerge at moderate $B_\parallel$. This superconducting state becomes energetically less favorable with increasing $B_\parallel$ both because of the continuous reduction of the order parameter, i.e., the superconducting gap of an Ising superconductor[34], and the increasing energy cost of spatially modulated supercurrents. Eventually, another dissipationless state, in which Cooper pairs carry a layer dependent momentum, has a lower free energy. This latter state appears with a uniform penetration of $B_\parallel$. The phase transition is thus accompanied by the melting of a Josephson vortex lattice. Due to the involvement of this melting, the jump in the order parameter is relatively small in magnitude (insets of Fig. 4c). This is another feature that distinguishes the phase transition here from that between a uniform superconductor and the Zeeman type FFLO state. There, the order parameter in the uniform superconducting state is strongly suppressed at the phase boundary. Finally, the finite momentum in the orbital FFLO state is carried by

Cooper pairs in one layer relative to those in the neighboring layers. Pairing within each layer has no net momentum shift. In this regard, it can be more robust to disorder scattering within the layers. Still, heavy disorder may break the superconductor into randomly distributed islands, which is reminiscent to a spatially modulated superconductor with Josephson vortices. This effect can suppress the orbital FFLO state. The comprehensive understanding of the microscopic mechanism helps guide further exploration of the orbital FFLO state in a wide pool of superconducting materials.

**Acknowledgements**

We thank Ignacio Gutierrez for helpful discussions. This work is financially supported by the Ministry of Science and Technology of China (2022YFA1403103); National Natural Science Foundation of China (Grants No. 12361141820, No. 12274249, No. 52388201). A. F. M. gratefully acknowledges the Swiss National Science Foundation for financial support. M. L. acknowledges financial support from the Swiss National Science Foundation through the Ambizione program. K.W. and T.T. acknowledge support from the JSPS KAKENHI (Grant Numbers 20H00354 and 23H02052) and World Premier International Research Center Initiative (WPI), MEXT, Japan.

**Author contributions**

D.Z. conceived the project; M. L. fabricated the samples; Z. C. and M. L. carried out the transport experiments with the assistance of Y. Z. and L. Z; Z. C., H. Y., D. Z., H. L., M. L. analyzed the data. H. Y. and H. L. carried out the theoretical calculations. K. W. and T. T. grew the h-BN crystals. D. Z., M. L., H. L. wrote the paper with the input from Z. C., A. F. M., and Q.-K. X.